\documentclass[twocolumns]{aa} % for a paper on 1 column  
\usepackage{graphicx}
\usepackage{txfonts}
\usepackage{hyperref}
\begin{document} 
\title{Discovery of an extended Horizontal Branch in the Large Magellanic Cloud globular cluster NGC1835
\thanks{Based on observations with the NASA/ESA HST,
obtained under program GO 16361 (PI: Ferraro). The Space Telescope Science Institute is operated
by AURA, Inc., under NASA contract NAS5-26555}}

   \author{Camilla Giusti
          \inst{1}
          \inst{2}
          \and
          Mario Cadelano
          \inst{1}
          \inst{2}
          \and
          Francesco R. Ferraro
          \inst{1}
          \inst{2}
          \and
          Barbara Lanzoni
          \inst{1}
          \inst{2}
          \and
          Cristina Pallanca
          \inst{1}
          \inst{2}
          \and
          Maurizio Salaris
          \inst{3}
          \and
          Emanuele Dalessandro
          \inst{2}
          \and
          Enrico Vesperini
          \inst{4}
          \and
          Alessio Mucciarelli
          \inst{1}
          \inst{2}
          }

\institute{Dipartimento di Fisica \& Astronomia, Universit\`a degli Studi di Bologna, via Gobetti 93/2, I-40129 Bologna, Italy\\
         \and
   INAF - Astrophysics and Space Science Observatory Bologna, Via Gobetti 93/3, 40129, Bologna, Italy \\
          \and 
   Astrophysics Research Institute, Liverpool John Moores University, Liverpool L3 5RF, UK \\
           \and 
   Dept. of Astronomy, Indiana University, Bloomington, IN 47401, USA\\
}

    %\date{May 2023}

% \abstract{}{}{}{}{} 
% 5 {} token are mandatory
 
  \abstract {We present a high angular resolution multi-wavelength study of
    the massive globular cluster NGC 1835 in the Large Magellanic
    Cloud.  Thanks to a combination of optical and near ultraviolet
    images acquired with the WFC3 on board the HST, we performed a
    detailed inspection of the stellar population in this stellar
    system adopting a ``UV-guided search'' to
    optimize the detection of relatively hot stars. This allowed us to discover a remarkably extended horizontal branch (HB),
    spanning more than 4.5 magnitudes in both magnitude and colour
    from the region redder than the instability strip, up to effective
    temperatures of 30,000 K, and including a large population of RR
    Lyrae (67 confirmed variables, and 52 new candidates).    
    This is the first time that such a feature has been detected in an
    extra-Galactic cluster, demonstrating that the physical conditions
    responsible for the formation of extended HBs are ubiquitous.
    The acquired dataset has been also used to redetermine the cluster
    distance modulus, reddening, and absolute age, yielding
    $(m-M)_0=18.58$, $E(B-V)=0.08$, and $t=12.5$ Gyr, respectively.  
  }

\keywords{}

\maketitle
%
%-------------------------------------------------------------------
\section{Introduction}
\label{sec:intro}
The Large Magellanic Cloud (LMC) is the most massive ($\sim 10^{11}
M_\odot$; \citealp{erkal+2019}) satellite of the the Milky Way. It
hosts a rich system of star clusters, including massive globular
clusters (GCs) with properties similar to those of the Milky Way, and
lower-mass stellar systems similar to the Galactic open clusters.  As
in the case of our Galaxy, the study of the LMC stellar systems
provides deep insights into the star formation history
\citep{olszewski+1996, olsen+1998, brocato+1996,
  mackey+2003,baumgardt+2013}, the chemical enrichment
\citep[e.g.,][]{hill+2000, pietrzynski+2000, ferraro+2006, mucciarelli+2010, glatt+2010,cadelano+2022a}, and the past merger
history (e.g., \citealt{mucciarelli+2021}) of the host galaxy. The LMC
GCs cover a metallicity range comparable to that sampled by Galactic
clusters, but show a much broader range of ages (from a few million, to
several billion years), thus providing the ideal laboratory to
empirically
calibrate the so-called ``red giant branch (RGB) and asymptotic giant
branch (AGB) Phase Transitions'' (see, e.g., \citealt{ferraro+1995,
  ferraro+2004, mucciarelli+2006}). These are two crucial events in a
star cluster life that are expected to induce significant changes in
the spectral energy distribution (SED) as a function of time. The
empirical calibration of theoretical SEDs is a mandatory step for the
proper interpretation of the spectra of unresolved galaxies through
cosmic time (see \citealt{maraston2005}).

To the same purpose, also the accurate characterization of the
horizontal branch (HB) morphology of stellar systems is extremely
important. In fact, it is well known that this can have a strong
impact on the integrated light of stellar populations, affecting their
colours and line indices \citep{lee+2002, schiavon+2004,
percival+2011, dalessandro+2012}. In particular, HBs
with extended blue tails imply the presence of very hot stars that,
in unresolved stellar systems, can mimick the existence of young
populations even in cases where star formation stopped several Gyr
ago.  Indeed, the so-called ``UV upturn'' or ``UV excess'' observed in
early-type galaxies is mainly explained as due to blue HB stars (e.g.,
\citealt{greggio+1990,dorman+1993, dorman+1995,brown2004}). In addition, peculiar populations
of ``slowly cooling white dwarfs'' have been recently identified in
GCs with extended blue HBs, while they are not observed in stellar
systems where the HB is restricted to the red (cold) region \citep{chen+2021,chen+2022,chen+2023a}. The link between the HB morphology and the
presence/lack of slowly cooling white dwarfs is due to the fact that,
because of their small mass, the bluest HB stars skip the asymptotic
giant branch phase and therefore keep a relatively massive residual hydrogen envelope around the degenerate carbon-oxygen core.  Hydrogen thermonuclear
burning in this residual envelope then acts as an extra-energy source
during the white dwarf phase, thus slowing down the evolution and
resulting in observable populations of ``slowly cooling white dwarfs''
\citep{chen+2021}. In spite of its astrophysical importance, we still have
an incomplete understanding of the physical origin of the HB
morphology (the so-called ``second parameter problem''; see e.g. \citealt{catelan2009,gratton+2010,milone+2014}) and it is therefore important to keep collecting
observational data and improving the theoretical models of this
crucial evolutionary phase.

The present paper is devoted to the characterization of the stellar population and the age of
the old LMC cluster NGC 1835. This investigation is part of a
project aimed at determining the structural parameters, chronological and dynamical age (i.e., the level of
dynamical evolution) of the oldest stellar clusters in the LMC
\citep{ferraro+2019, lanzoni+2019}. In particular the dynamical age is determined by using the so-called ``dynamical
clock'' that quantifies the central segregation of blue straggler
stars (see \citealt{ferraro+2012, ferraro+2018, ferraro+2020,
  ferraro+2023, lanzoni+2016}).
As part of this project, we secured a set of multiband Hubble Space
Telescope (HST) images of NGC 1835, a very massive ($\sim6\times10^5
M_\odot$) and old ($t\sim13$ Gyr) globular cluster lying close to the
central bar of the LMC \citep{mackey+2003}. The necessity of detecting
faint blue objects, like blue straggler stars, in the central regions
of this high-density and distant cluster imposed the use of an
efficient filter (F300X) with sensitivity in the near ultraviolet
(near UV) domain. The observations performed with this new eye have
revealed an unexpected feature: the HB of NGC 1835 shows a very
extended blue tail. This remained undetected in all previous optical
studies of the cluster (see, e.g., \citealp{olsen+1998}) and
represents the first evidence ever collected in an extra-Galactic
cluster. 
These images have already provided evidence of the presence of the small stellar system KMK 88-10, which appears to be in the process of being gravitationally captured by the massive GC NGC 1835 \citep{giusti+2023}. In addition, while a forthcoming paper will be devoted to the
determination of the dynamical age of NGC 1835 (Giusti et al., in
preparation), the present work is focused on the determination of the chronological age and the characterization of the HB morphology of the
cluster.

The paper is structured as follows. In Section \ref{sec:reduction} we
describe the data set and the data reduction process. In Section
\ref{sec:cmd} we present the main features of the color-magnitude
diagram (CMD).  Section \ref{sec:HB} presents the main characteristics
of the HB in NGC 1835, discussing its population of RR Lyrae variable
stars (Section \ref{sec:RRlyrae}) and the decontamination from LMC
field stars (Section \ref{sec:HBdeco}).  Section \ref{sec:params} is
devoted to the determination of the reddening, distance modulus, and
age of the cluster, which are then needed for a deeper discussion of
the surprising HB morphology through the comparison with two reference
Galactic GCs and the determination of the effective temperature
distribution of HB stars (Section \ref{sec:HBmorph}).  The summary and
discussion of the results are provided in Section \ref{sec:conclu}.

%%%%%%%%%%%%%%%%%%%%%%%%%%%%%%%%%%%%%%%%%%%%%%%%%%%%%%%%%%%%%%%%%%%%%%%%%%%%%
\section{Data set and data reduction}
\label{sec:reduction}
The photometric study of NGC 1835 was performed using a data set
of deep and high-resolution images obtained with the UVIS channel of
the Wide Field Camera 3 (WFC3/UVIS) onboard the HST (program GO 16361,
PI: Ferraro). A total of 16 images were acquired using the F300X,
F606W and F814W filters to cover a broad range of
wavelengths. Specifically, 6 images ($2\times$ 900 s, $1\times$ 917 s,
$2\times$ 920 s, $1\times$ 953 s) were taken in the near UV F300X
filter, 6 images ($2\times$ 407 s, $4\times$ 408 s) were acquired in
the F606W, and 4 exposures ($1\times$ 630 s, $1\times$ 645 s,
$2\times$ 700 s) in the F814W. In each pointing, the center of NGC
1835 was aligned with the center of the WFC3 UVIS1, while UVIS2
sampled distances out to approximately $120\arcsec$.  The
WFC3 data set was complemented with simultaneous parallel
observations
acquired with the Wide Field Camera of the Advanced Camera for Surveys
(ACS/WFC) in the F606W and F814W filters. These cover a
$200\arcsec\times 200\arcsec$ wide region located at $\sim 5\arcmin$
from the cluster, thus properly sampling the LMC field contaminating
the cluster population.  A total of 7 images ($1\times$ 335 s,
$3\times$ 340 s, and $3\times$ 350 s) were acquired in the F606W
filter, and 6 ($1\times$ 550 s, $5\times$ 600 s) in the F814W .

We have performed the data reduction using the software DAOPHOT II
\citep{stetson+1987}, following the recipes described in detail in
\citet{cadelano+2020b, cadelano+2020c, deras+2023, deras+2024}.  We selected about 200 bright,
well-distributed, and isolated stars in order to model a spatially
varying point spread function (PSF) for each image.
The PSF model thus obtained was then used to perform the PSF-fitting
of all the sources with a flux peak above 5$\sigma$ from the
background level. We created a reference master list of stars
identified in at least half of the images acquired with the UV filter,
and we then forced the fit of the PSF model to the location of these
sources in all the other images, using DAOPHOT/ALLFRAME
\citep{stetson+1994}.  This approach is called the ``UV-guided
search'', and it was proposed and extensively adopted by our group
(see \citealt{ferraro+1997a, ferraro+1998, ferraro+1999a,
  ferraro+2001, ferraro+2003, lanzoni+2007a,dalessandro+2013}, and
more recently \citealt{raso+2017,chen+2021,chen+2022,chen+2023b,cadelano+2022b}) to optimize the
detection and obtain complete samples of hot stars (such as
extremely-blue HB stars, blue straggler stars, and white dwarfs) in
stellar populations dominated by cool giants, like old star clusters.
In fact, the technique mitigates the crowding effects caused by the
presence of giants and main sequence turn-off (MS-TO) stars, making it
easier to retrieve blue and hot stars that would be lost in optical
and infrared images, thus sensibly increasing the level of
completeness of the observed samples.

For each identified star, the magnitudes estimated in different images
were combined using DAOMATCH and DAOMASTER. The final catalogues
consisted of frame coordinates, instrumental magnitudes, and
photometric errors for a total of more than 100,000 sources:
approximately 65000 sources in the WFC3 catalogue sampling the entire
cluster extension, and 82000 in the ACS catalog, sampling the LMC
field. The magnitudes have been calibrated onto the VEGAMAG
photometric system by applying appropriate aperture corrections and
the zero points reported on the HST WFC3 and ACS
websites.
Finally, after the geometric distortion effects were corrected by
applying the coefficients from \citet{bellini+2011} for WFC3 and from \citet{meurer+2003} for ACS, the positions
were transformed to the absolute coordinate system ($\alpha, \delta$)
by cross-correlation with the Gaia DR3 catalogue
\citep{gaia+2022} sampling the same regions of the
  sky.

\begin{figure*}[ht!]
    \centering
    \includegraphics[scale=0.35]{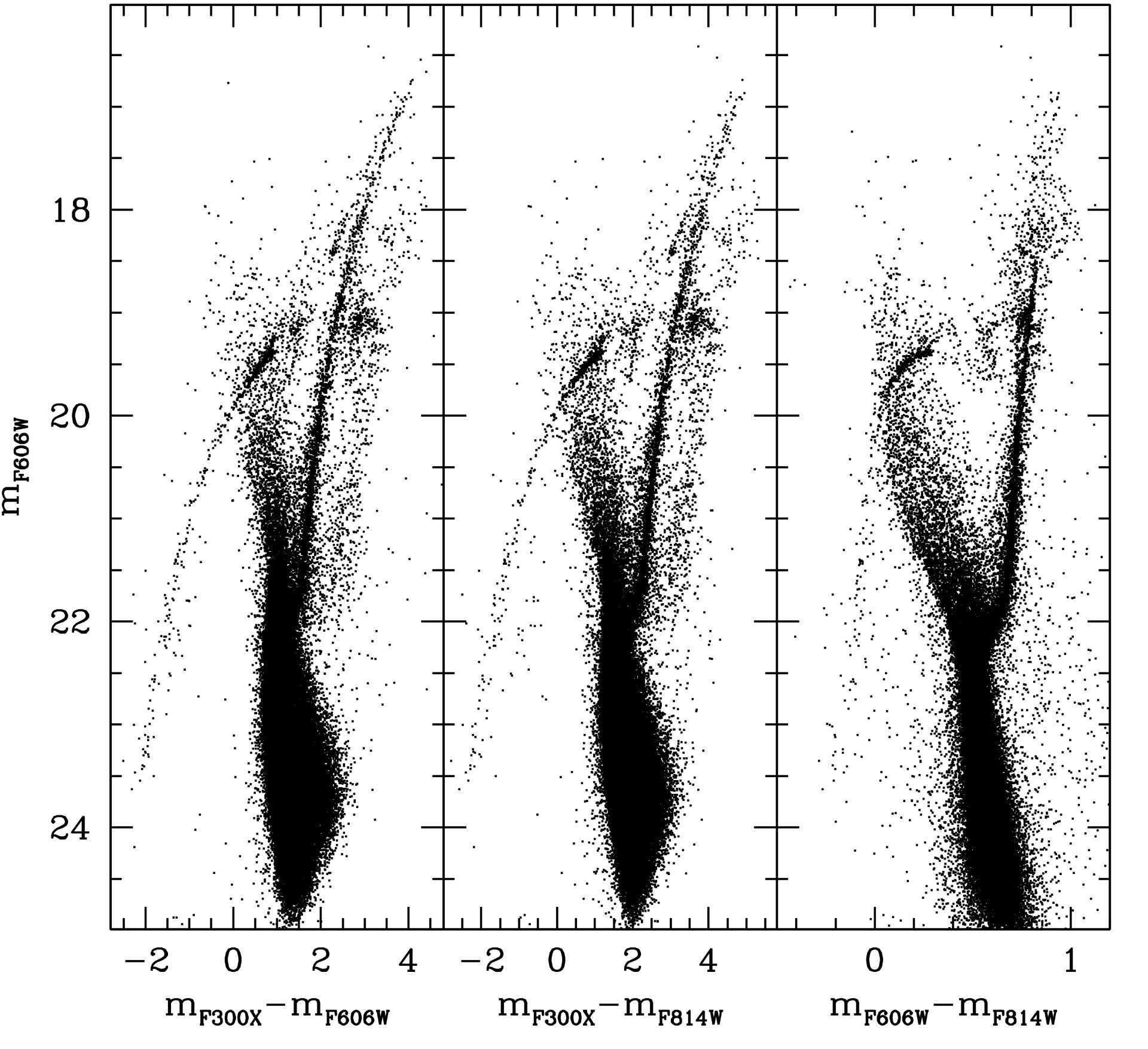}
    \caption{CMD of NGC 1835 obtained from the WFC3 data set in all
      the filters combination.  }
    \label{fig:cmds}
\end{figure*}

\begin{figure}[ht!]
    \centering
    \includegraphics[scale=0.3]{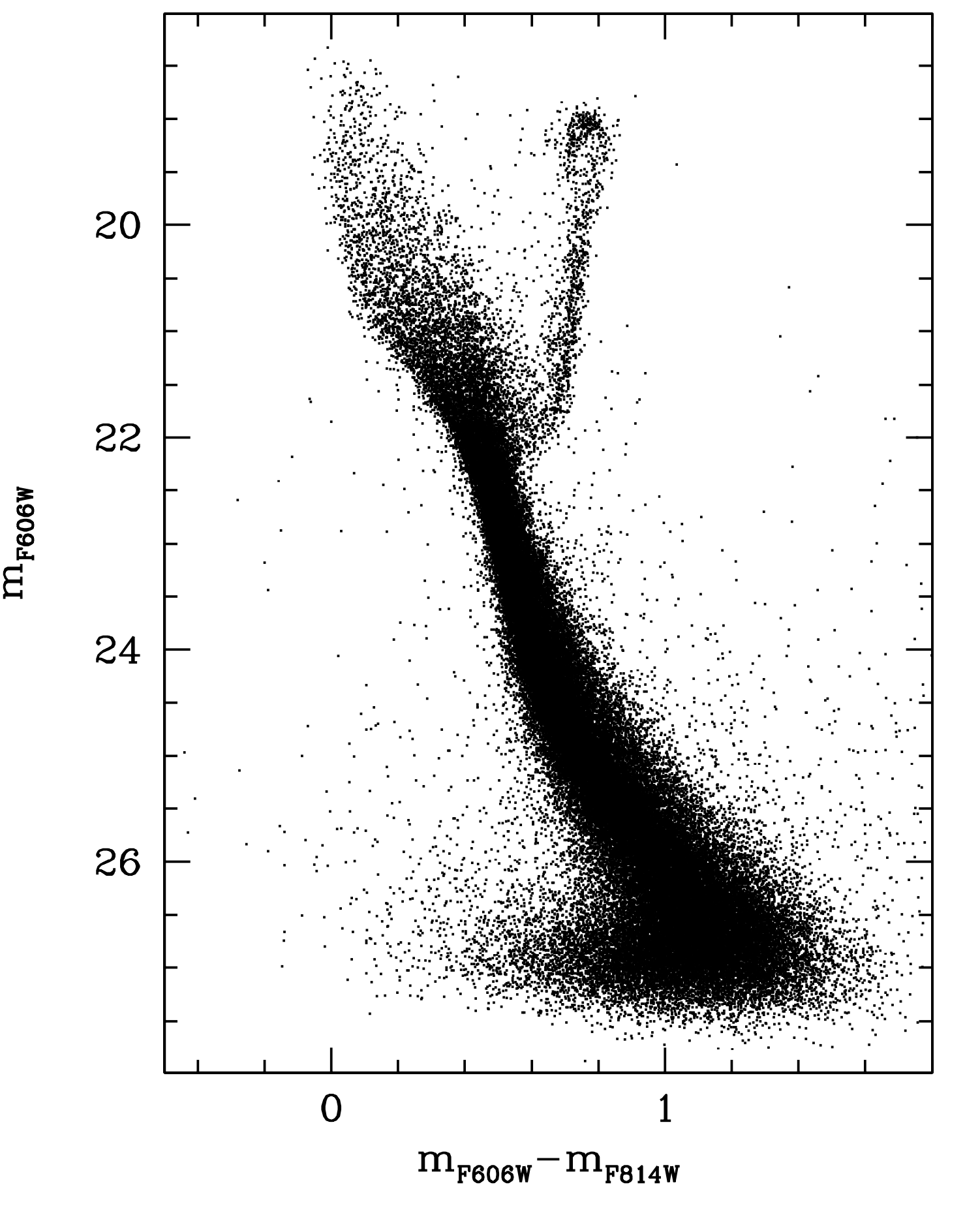}
    \caption{CMD of the LMC field obtained from the ACS parallel
      observations.}
    \label{fig:cmd_acs}
\end{figure}

%%%%%%%%%%%%%%%%%%%%%%%%%%%%%%%%%%%%%%%%%%%%%%%%%%%%%%%%%%%%%%%%%%%%%%%%%%%%%
\section{The colour-magnitude diagram}
\label{sec:cmd}
Figure \ref{fig:cmds} shows the CMDs obtained from the WFC3
observations in all the filter combinations. The optical CMD of the
LMC field sampled by the ACS observations is shown in Figure
\ref{fig:cmd_acs}.  From a visual analysis of these figures it appears
clear that the CMD of NGC 1835 is significantly contaminated by field
stars, in agreement with what expected for a GC located close to the
central bar of the LMC. However, the main evolutionary sequences
belonging to NGC 1835 are clearly distinguishable in the CMD. The most
severe contamination is produced by the MS of the LMC field, which is
essentially superposed to the cluster MS and extends up to magnitudes
brighter than the cluster HB (up to $m_{\rm F606W} \sim 18$). The RGB
of the LMC field is clearly distinguishable from that of NGC 1835 when
colours including the F300X filter are used (left and central panels
of Fig. \ref{fig:cmds}). In these colour combinations also the field
Red Clump is well visible (the rightmost clump of stars at $m_{\rm
  F606W}\sim 19$).  On the other hand, no significant field
contamination is observed in the bluest region of the CMD, at colours
bluer than 0.  This is the region where the most intriguing and
unexpected feature of the CMD (i.e., a remarkably extended blue tail of the
HB; see below) is located.

%%%%%%%%%%%%%%%%%%%%%%%%%%%%%%%%%%%%%%%%%%%%%%%%%%%%%%%%%%%%%%%%%%%%%%%%%%%%%
\section{The horizontal branch of NGC 1835}
\label{sec:HB}
The HB extension of NGC 1835 is fully appreciable in the CMDs
involving the F300X filter, which has ``guided'' the detection of hot
stars (see Sect. \ref{sec:reduction}).
However, the peculiarity of this HB is not only its magnitude
extension, but also its color distribution: in fact, the HB appears
to be well populated both in its red ($0.5 < m_{\rm F300X}-m_{\rm
  F606W} < 0.8$) and blue ($-2.1 < m_{\rm F300X}-m_{\rm F606W} <0.4$)
portions, and also shows a large population of stars in the
instability strip.

%%%%====================================================================
\subsection{The RRLyrae population}
\label{sec:RRlyrae}
The identification of RR Lyrae variable stars is a crucial step for
the proper analysis of the HB morphology.  For their identification we
exploited a photometric ``variability index'', defined as the ratio
between the standard deviation of the star luminosities measured at
different epochs (which provides an estimate of the amplitude of the
light curve) and the internal photometric error (which provides an
estimate of the quality of the star's photometry; \citet{kjeldsen+1992}. For stars having magnitude values from different
exposures that consistently fall within the photometric error, the
variability index tends to approach unity. Conversely, an increase in
the index value corresponds to an increasing probability of stellar
variability.  In the case of NGC 1835,  we found that the variability
  indexes computed in the near UV tend to be larger than those
  obtained in the optical bands, thus suggesting that the F300X filter
  is best suited for this kind of study. We therefore assumed as
  variable stars all the sources with a variability index larger than
  2 in this filter.

\begin{figure}[ht!]
    \centering
    \includegraphics[scale=0.5]{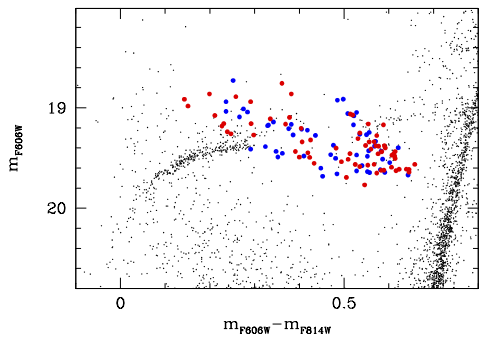}
    \caption{CMD of NGC 1835 zoomed in the HB region showing the  67 confirmed OGLE variables, and the 52 new candidate RRLyrae are highlighted as red and blue circles, respectively. Only non-variable stars observed at $r<60\arcsec$ from the centre are shown for reference as black dots.}
    \label{fig:rrlyrae}
\end{figure}

The number of variable stars recovered with this approach is
95. However, since our dataset was not originally intended for
variability studies, we complemented this information with the OGLE
IV catalogue\footnote{https://ogledb.astrouw.edu.pl/~ogle/CVS/}, which provides information on variable stars and
transient objects in the Magellanic Clouds and Galactic
bulge. This catalogue presents an updated list of variables found in the direction of the cluster NGC 1835, for a total of 125 objects, mainly RR Lyrae \citep{soszynski+2016}.
After proper cross-correlation, we found a total of 67 variable stars in the region of the instability strip in common between our catalogue and the OGLE one. Interestingly enough, 43 of them
were also identified via the variability index, while the remaining 24
OGLE variables were possibly caught in low luminosity variation
phases by our observations. In addition, the DAOPHOT variability index indicates that in the field of view sampled by our observations there are 52 more candidate RR Lyrae that are not included in the OGLE catalogue.
Thus, NGC 1835 hosts a very large population of RR Lyrae variables, counting 119 objects in the field of view sampled by our observations: 67 are confirmed OGLE variables, and 52 are possible new candidates identified
by the variability index (see Figure \ref{fig:rrlyrae}).

%%%%====================================================================
\subsection{Field decontamination and extension of the  HB}
\label{sec:HBdeco}
As can be appreciated in Fig. \ref{fig:cmds},
the main LMC field contamination along the HB is in the red portion of
the branch, in the colour range $0<m_{\rm F300X}-m_{\rm F606W} < 1.2$,
with a minor contribution in the adjacent colour bin $1.2<m_{\rm
  F300X}-m_{\rm F606W} < 2$ (``extremely-red'' portion). With the available datasets
(only HST/WFC2 images of the cluster are present in the HST archive),
the decontamination through proper motions is not feasible. In fact,
given the large distance of the LMC \citep{harris+2009,
  pietrzynski+2019},
a first epoch dataset with an astrometric quality better than that
provided by the archive WFPC2 observations is required, possibly
combined with a significantly longer time baseline between the two
epochs. Thus, we
performed a statistical decontamination of these portions of the HB
taking advantage of the parallel ACS observations \citep{ferraro+2019, dalessandro+2019}. Schematically,
after removing all the confirmed and candidate variables, we selected
the HB stars in the $m_{\rm F606W}, m_{\rm F300X}-m_{\rm F606W}$
CMD. Then, we used the position of these stars in the $m_{\rm F606W},
m_{\rm F606W}-m_{\rm F814W}$ CMD to draw the \emph{optical} selection
boxes of the red and extremely-red HB portions. We counted the number
of stars belonging to ACS catalogue that fall within the two selection
boxes and, dividing by the area sampled by the parallel observations,
we finally obtained the density of LMC field stars contaminating the
red and the extremely-red portions of the HB.  This procedure yielded
a field density contamination of 17.28 and 3.44 stars per square
arcminute, which corresponds to a total of 120 and 25 field stars
potentially contaminating the red and the extremely-red HB portions,
respectively.  The statistical decontamination was then performed
taking into account the distance from the cluster center and the
geometric distance of each stars from the mean ridge line of the HB,
since the membership probability typically decreases if these
distances increase.  Hence, the cluster (WFC3) sample has been divided
in concentric annuli, and the expected number of contaminating stars
was determined in each radial annulus from the estimated field density
and the bin area. Finally, the estimated number of field stars has
been removed from the WFC3 sample starting from the objects located at
larger geometric distances from the HB mean ridge line.

\begin{figure}[ht!]
    \centering
    \includegraphics[scale=0.25]{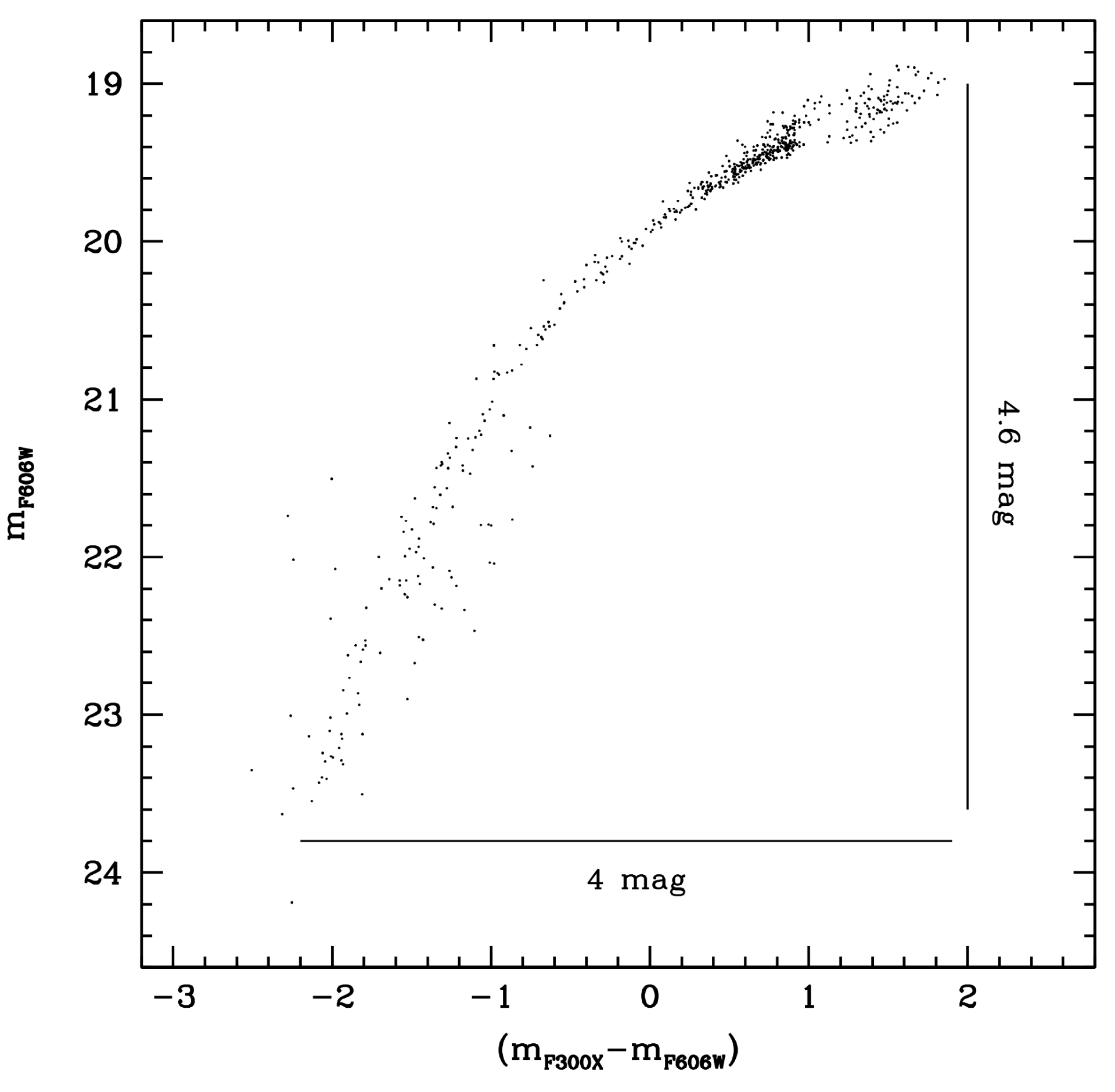}
    \caption{HB of NGC 1835 after the statistical decontamination of
      the reddest portions. Only non-variable stars are plotted. The
      extension in colour ($\sim$4 mag) and in magnitude ($\sim$ 4.6
      mag) is also marked.}
    \label{fig:hb}
\end{figure}

The HB of NGC 1835 obtained after statistical decontamination counts
602 non-variable stars and is plotted in Figure \ref{fig:hb}.  For the
sake of clarity, we also explicitly mark the large extension in colour
($\sim 4$ mag) and in magnitude ($\sim 4.6$) of the branch. Indeed,
even neglecting the faintest stars detected at $m_{\rm F606W}\sim
24.2$, the extension in magnitude and colour is remarkable: this is
the first time that such an extended HB is detected in an
extra-Galactic cluster.

%%%%%%%%%%%%%%%%%%%%%%%%%%%%%%%%%%%%%%%%%%%%%%%%%%%%%%%%%%%%%%%%%%%%%%%%%%%%%
%\subsection{Comparison with M3 and M13}
%\label{sec:m3m13}
\section{Reddening, distance modulus and age of NGC 1835}
\label{sec:params}
For a more detailed description of the HB morphology of NGC 1835 we
considered two Galactic GCs (namely, M3 and M13) as reference
systems. These clusters have been extensively studied in the
literature and their properties are well established (see
\citealt{ferraro+1997b, johnson+2005, sneden+2004,
  dalessandro+2013,cadelano+2019,cadelano+2020a}). They are two ``twin'' GCs, with similar age,
metallicity, and central density. They have respectively an age of
$12.50 \pm 0.50$ Gyr and $13.00 \pm 0.50$ Gyr \citep{dotter+2010}, [Fe/H] = $-1.50 \pm 0.05$ and [Fe/H] $= -1.58 \pm 0.04$
\citep{carretta+2009}, and a $V-$band central surface brightness
  $\mu_0=16.64$ mag arcsec$^{-2}$ and 16.59 mag arcsec$^{-2}$\citep{harris+1996}. In addition, they have a similar mass of about $\log M/M_\odot \sim 5.8$ \citep{mclaughlin+2006,pryor+1993}. These values are
compatible with those measured in NGC 1835: [Fe/H]$=-1.69\pm0.01$ \citep{mucciarelli+2021}, $\mu_0=16.64$ mag arcsec$^{-2}$ in the F555W filter \citep{mackey+2003} and $\log M/M_\odot \sim 5.83$ \citep{mackey+2003}. As for the age, there is general agreement on the fact that NGC 1835 belongs to the class of old LMC clusters, but the absolute age is still quite uncertain and strongly overestimated: $\log
t =10.2^{+0.07}_{-0.08}$, corresponding to an age $t \sim$ 16 Gyr (see \citep{olsen+1998, olszewski+1991}). In addition, the HB of M13 shows a well extended and populated blue tail, reminiscent of that discovered in NGC 1835, but very few RR Lyrae (only 10 are quoted in the Clement catalog; \citealp{clement+2001}\footnote{https://www.astro.utoronto.ca/\textasciitilde cclement/read.html}) and just a few stars redder than the instability strip. Conversely, M3 hosts a large population of RR
Lyrae (more than 200), similarly to NGC 1835, but no evidence of the blue tail. Hence, M3 and M13 are excellent reference clusters for an in-depth study
of NGC 1835. Indeed, through the direct comparison of the CMDs we
determined the reddening and the distance of NGC 1835, which are then
used to estimate the age of the cluster. With this information in hand, we finally proceeded
to a direct comparison between the HBs of the three systems, as well
as the determination of the effective temperature distribution of the
HB stars in NGC 1835 (Section \ref{sec:HBmorph}).

%%..............
%\subsection{Reddening, distance modulus and age of NGC 1835}
%\label{sec:params}
We performed the cluster-to-cluster comparison in the
optical $(m_{\rm F606W}, m_{\rm F606W}-m_{\rm F814W})$ CMD. To maximize the accuracy of such a comparison, we first constructed a
field-decontaminated CMD of NGC 1835. We decided to perform a statistical
decontamination following the prescriptions described in
\citet{giusti+2023} and briefly summarized below.
%In doing this we decide to apply
We considered an annular region of the CMD included between $5\arcsec$
and $26\arcsec$ from the cluster center.  The very central region has
been excluded because it is typically sampled with larger photometric
errors, and the most external one is rejected to keep low the number
of contaminating field stars. Then, a region with the same area has been selected in the ACS
catalogue of the LMC field and, for each star of this sample, we
removed one star from the cluster sample according to its position in
the CMD. In particular, we flagged as the most likely interloper (and
removed it from the cluster sample) the closest star (in terms of
magnitude and colour) to the considered star in the LMC field
sample. The procedure was repeated several times considering different
LMC field regions, finding qualitatively consistent results in all the
cases. 

\begin{figure*}[ht!]
    \centering
    \includegraphics[scale=0.3]{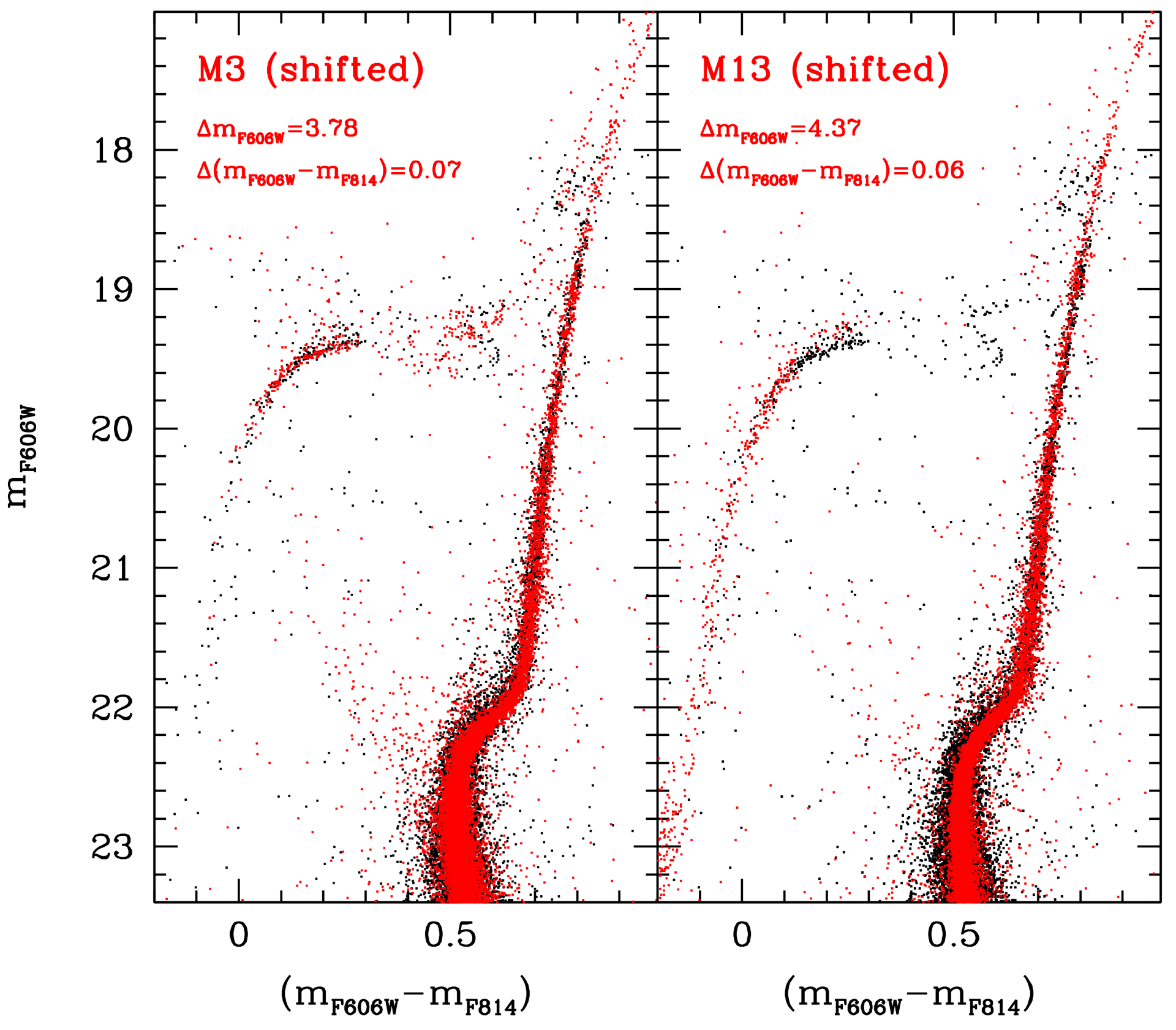}
    \caption{Comparison between the decontaminated CMD of NGC 1835
      (black dots) and the CMDs of M3 and M13 shifted to match the
      former (red dots in the left and right panels,
      respectively). }
\label{fig:cmd_confronto}
\end{figure*}

The field-decontaminated CMD of NGC 1835 thus obtained was then compared with
those of the two reference clusters, M3 and M13, by using the data collected in the
\textit{ACS Survey of Galactic Globular Clusters}
\citep{anderson+2008}. We then aligned these CMDs with that of NGC
1835 by applying appropriate shifts in magnitude and colour, chosen on
the basis of $\chi ^2$ tests: for M3 we used $\Delta_{\rm mag} = 3.78$
and $\Delta_{\rm col} = 0.07$; for M13 we applied $\Delta_{\rm mag} =
4.37$ and $\Delta_{\rm col} = 0.06$. The left panel of Figure
\ref{fig:cmd_confronto} presents the comparison between the
decontaminated CMD of NGC 1835 (black dots) and the CMD of M3 (red
dots). The right panel shows the same, but for M13. Most of the
evolutionary sequences of NGC 1835, such as the MS and the RGB,
exhibit striking similarities with those of M3 and M13. Specifically,
all the three clusters show a comparable morphology of the MS-TO
region and have a similar colour extensions of the sub-giant
branch. Moreover, the RGBs of the three clusters display comparable
extensions and slopes. These results are consistent with the
three clusters having similar ages and metallicities, as discussed
above.

Using the extinction coefficients $R_{\rm F606W}=2.8192$ and $R_{\rm
  F814W}=1.8552$ \citep{cardelli+1989, odonnell+1994} and assuming for
M3 a distance modulus of $(m-M)_0 = 15.00\pm 0.04$
\citep{ferraro+1999b, dalessandro+2013} and a colour excess
$E(B-V)=0.01$ \citep{ferraro+1999b}, we found $(m-M)_0 = 18.58$ and
$E(B-V) = 0.08$ for NGC 1835. The same results
are obtained by adopting M13 as a reference and thus assuming its
distance modulus and reddening, $(m-M)_0 = 14.38\pm 0.05$ and
$E(B-V)= 0.02$ \citep{ferraro+1999b}. These findings are fully
consistent with previous estimates in the literature: in fact, a
reddening value of $E(B-V)=0.08 \pm 0.02$ and a distance modulus
ranging between $18.43 \pm 0.12$ and $18.65 \pm 0.16$ are quoted in
\citet{olsen+1998}.

%%..............
To determine the absolute age of NGC 1835 we adopted the isochrone
fitting technique. This consists in comparing the cluster CMD with a
set of isochrones of different ages to identify the one that best
reproduces the observed evolutionary sequences. We extracted
isochrones from two different databases, BaSTI
\citep{pietrinferni+2021} and PARSEC \citep{bressan+2012, chen+2015}.  BaSTI isochrones have been
computed for [Fe/H]$= -1.7$, an $\alpha-$element
abundance [$\alpha$/Fe]$= +0.4$ \citep{mucciarelli+2021}, and a
standard helium abundance Y = 0.248. Similarly, we downloaded solar scaled PARSEC isochrones with a total metallicity [M/H] = $= -1.4$, that corresponds to [$\alpha$/Fe]$= +0.4$ and [Fe/H]$= -1.7$. Both datasets were downloaded for a suitable
range of ages (between 10 Gyr and 15 Gyr in steps of 0.5 Gyr). They have been superposed
to the observed CMD by adopting the distance modulus and the reddening
values estimated above. A small color offset of $\delta(m_{\rm F606W}-m_{\rm F814W})=0.015$ mag (lower than the typical errors introduced by the photometric calibration) was applied to the isochrone in order to better match the data. Using $\chi^2$ tests, we compared the
isochrones and the observed data in the most age-sensitive region of
the CMD (namely, the MS-TO and the sub-giant branch, in a magnitude
range $21.8 < m_{\rm F606W} < 23.0$). For a maximum photometric
quality of the data and to avoid severe contamination from the LMC
field, we limited the analysis to an annular region between
$15\arcsec$ and $26\arcsec$ from the cluster center.

\begin{figure*}[ht!]
    \centering
    \includegraphics[scale=0.3]{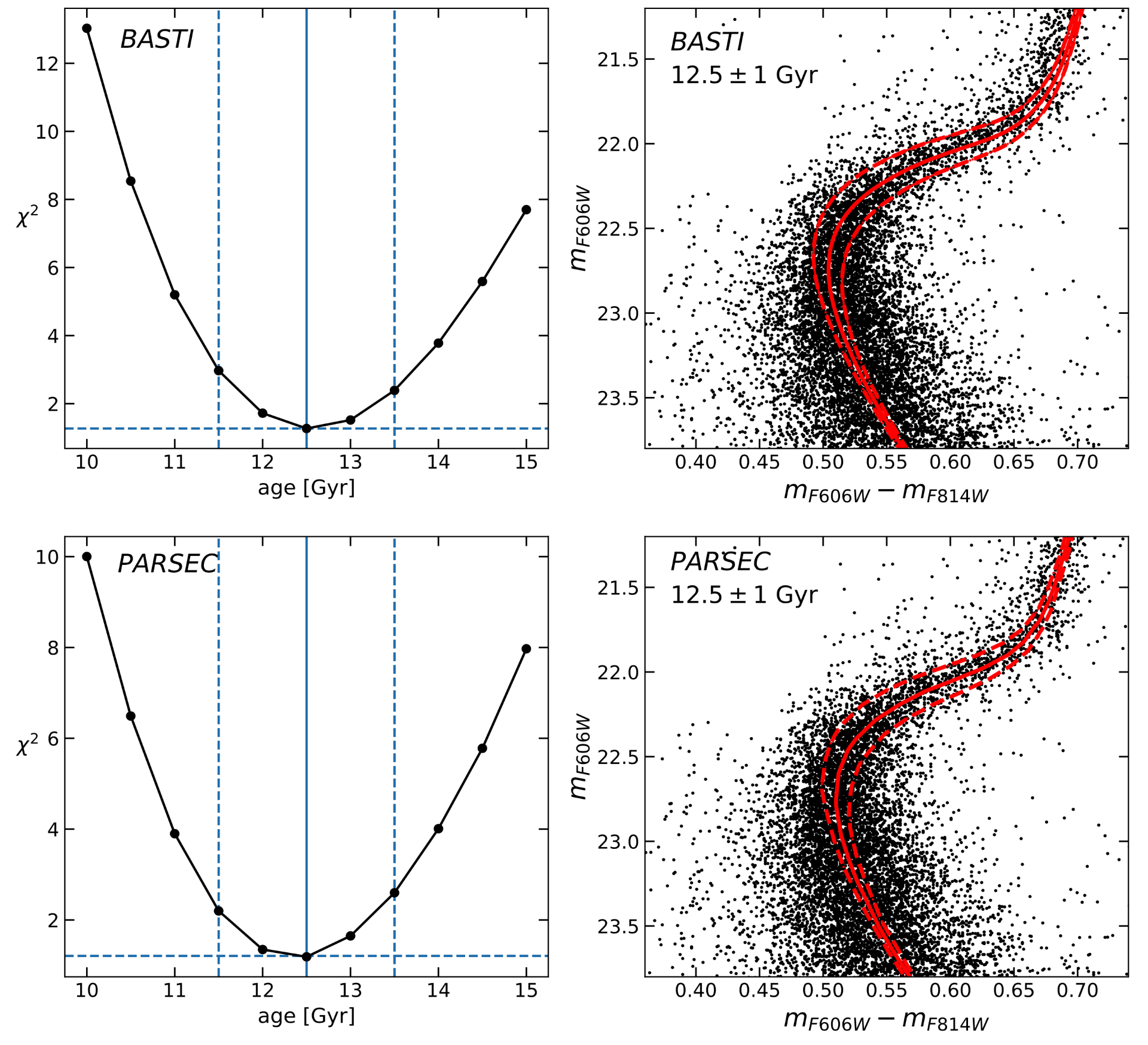}
    \caption{\emph{Left panels:} Trend of the $\chi^2$ values
        obtained from the isochrone fitting of the MS-TO region of the
        models (top and bottom panels, respectively) computed for
        [Fe/H]$= -1.7$, [$\alpha$/Fe]$= +0.4$, Y = 0.248 and ages
        ranging between 10 Gyr and 15 Gyr.  \emph{Right panels:}
        decontaminated CMD of NGC 1835 with superposed the BaSTI and
        PARSEC isochrones (top and bottom panels, respectively)
        corresponding to the best-fit age (12.5 Gyr, solid lines) and
        the estimated age interval ($\pm 1$ Gyr, dashed lines).   }
\label{fig:age}
\end{figure*}

The results obtained for the two considered models are shown in Figure
\ref{fig:age}.  The left panels depict the trend of the $\chi^2$
  values for different ages and show that, for both the adopted model,
  the minimum $\chi^2$ value (i.e., the best-fit solution) is found at
  an age of 12.5 Gyr, with a conservative uncertainty of $\pm 1$
  Gyr. The right panels of Fig. \ref{fig:age}
  show the isochrones corresponding to the best-fit ages (solid red
  line) along with the uncertainty ranges (dashed lines).
 
%%%%%%%%%%%%%%%%%%%%%%%%%%%%%%%%%%%%%%%%%%%%%%%%%%%%%%%%%%%%%%%%%%%%%%%%%%%%%
\section{The surprising HB morphology of NGC 1835}
\label{sec:HBmorph}
As mentioned above, the HB of M13 shows a well extended and populated
blue tail, very few RR Lyrae, and the portion redder than the
instability strip that is poorly populated.  Conversely, in M3 the
star distribution is peaked around the instability strip, with a large
population of RR Lyrae, and the red and the blue side of the strip are
both well populated, but with no evidence of blue tail. These
characteristics make the two Galactic GCs particularly useful for the
description of the HB morphology in NGC 1835.

\begin{figure}[ht!]
    \centering
    \includegraphics[scale=0.25]{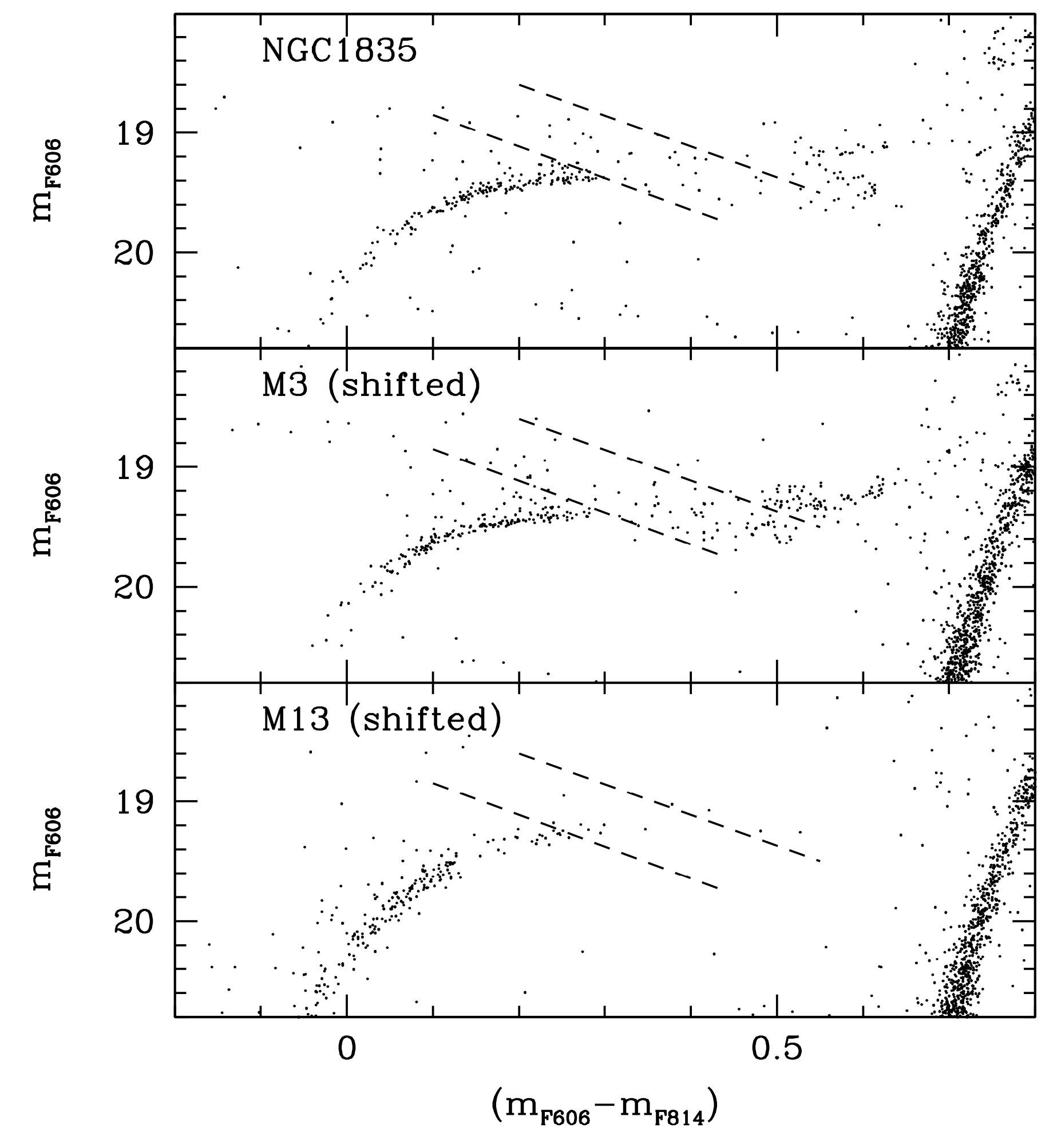}
    \caption{Low-temperature portion of the HB of NGC 1835 (top panel)
      compared with that of M3 (central panel) and M13 (bottom panel). The CMD of M3 and M13 have been
      shifted to the same distance and reddening of NGC 1835. In all
      panels, the two dashed lines delimitate the region where most of the RR
      Lyrae observed at random phases are expected to be located.}
    \label{fig:hbred}
\end{figure}

As shown in Fig.\ref{fig:hb}, the colours using the F300X filter tend to stretch the
high-temperature tail of the branch, while compressing the
low-temperature portion. Thus, a better insights into the star distribution in
the red (cold) side of the HB are obtainable in the purely optical
colour ($m_{\rm F606W} - m_{\rm F814W}$), which also allows a clear
identification of the location of the instability strip.  Fig.
\ref{fig:hbred} shows the comparison in the optical CMD among the
low-temperature portion of the HB observed in NGC 1835 (top panel),
in M3 (central panel) and in M13 (bottom panel).  The CMD of M3 and M13 have been shifted to the distance
and the reddening of NGC 1835 estimated above. The similarity between M3 and NGC1835 in the morphology of this portion of the HB is
impressive, and it is further supported by the existence of large
populations of RR Lyrae detected in both systems. While it is well evident that the HB in M13
extends mainly on the blue side of the instability strip, hence only a few (possibly evolved HB stars) 
are expected to appear as RRLyrae variable, as confirmed by the low number 
of confirmed variables in this cluster.

\begin{figure}[ht!]
    \centering
    \includegraphics[scale=0.25]{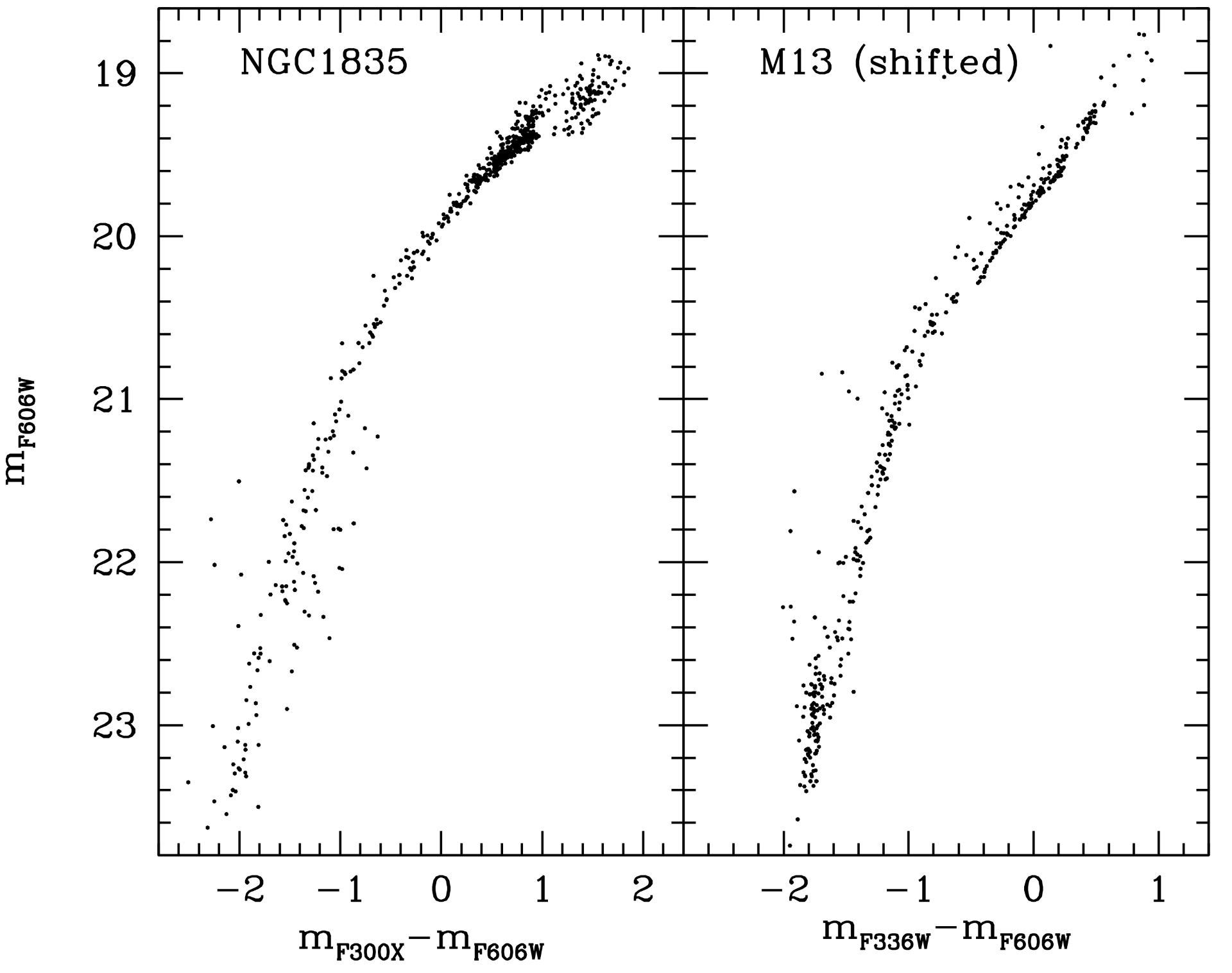}
    \caption{HB of NGC 1835 (left panel) compared to that of M13
      (right panel). Variable stars are not shown.  }
    \label{fig:tail}
\end{figure}

For the sake of comparison,
Figure \ref{fig:tail} shows instead the comparison between the HB of NGC 1835
(left panel) and that of M13 (right panel) in the optical-near UV hybrid
CMD. As usual the CMD of M13 has been shifted to the distance
  of NGC 1835 estimated above.  Only non-variable stars are plotted here. Although the filters
used in this comparison are not exactly the same, the overall
magnitude extension of the branch is strikingly similar. Notably, the
colour extension in NGC 1835 looks wider than in M13, with several
stars reaching redder colours at the right-end side of the instability
strip, where almost no objects are observed in M13 (see also Fig.\ref{fig:hbred}).
 
This comparison therefore shows that the HB observed in NGC 1835 looks
like a combination of the HBs of M3 and M13, with a well populated red
portion characterized by a relevant number of RR Lyrae (like in M3,
and at odds with M13), and presenting an extended blue tail (at odds
with M3, but as in M13).
 
\begin{figure*}[ht!]
    \centering
    \includegraphics[scale=0.6]{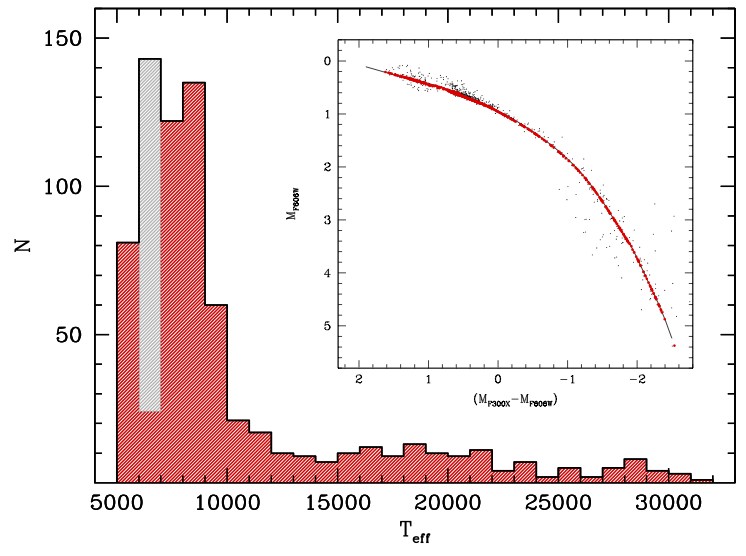}
    \caption{Effective temperature distribution of the HB stars in NGC
      1835. The histogram shaded in gray refers to the sample of
      119 confirmed and candidate variable stars
      distributed within the range of temperatures covered by the
      instability strip ($6000 < T_{\rm eff} < 7000$). The inset shows
      the locations of non-variable HB stars in the absolute CMD
      (black dots), and the corresponding positions projected onto the
      HB mean ridge line (red dots along the black solid line).  The
      colour axis is inverted (with bluer colors to the right) to be
      consistent with the x-axis of the histogram (increasing
      temperature to the right). }
\label{fig:teff}
\end{figure*}

As last piece of information, we determined the temperature
distribution of the HB stars in NGC 1835. In doing this, we first
translated the decontaminated HB into the absolute plane, adopting the
distance modulus and the reddening determined in Section
\ref{sec:params}.  We then projected the non-variable stars perpendicularly onto the
mean ridge line of the branch (see the inset in Figure
\ref{fig:teff}), thus obtaining their projected intrinsic colour and
absolute magnitude.  Then, to transform the ($M_{\rm F300X}-M_{\rm
  F606W}$) colour into effective temperature ($T_{\rm eff}$), we
adopted the colour-temperature relation for zero-age HB models of
appropriate metallicity ([Fe/H]$=-1.7$) extracted from the BASTI
dataset \citep{pietrinferni+2021}.  To complete the overall picture,
we arbitrarily attributed $T_{\rm eff}$ values to all the confirmed
and candidates RR Lyrae in the temperature range covered by the
instability strip ($6000< T_{\rm eff} < 7000$).  The resulting
$T_{\rm eff}$ distribution (see Fig. \ref{fig:teff}) shows a main peak
in the red portion of the HB (possibly in the instability strip), and
a long tail extending to temperatures as large as 30,000 K. The tail
is not uniformly populated (in agreement with what observed in all
other similar cases; see, e.g., the discussion in
\citealt{ferraro+1998}), but it seems to show a secondary peak (by far
less populated than the main one) in the region between 15,000 and
20,000 K.
  
%%%%%%%%%%%%%%%%%%%%%%%%%%%%%%%%%%%%%%%%%%%%%%%%%%%%%%%%%%%%%%%%%%%%%%%%%%%%%
\section{Summary and Conclusions}
\label{sec:conclu}
In the framework of a project aimed at performing a new
characterization of the oldest and most compact stellar systems in the
LMC, we have presented a detailed multi-wavelength photometric study
of the GC NGC 1835. A set of near UV and optical high-resolution
images have been secured using the HST/WFC3. These images already
revealed the presence of the small stellar system KMK 88-10 at just
$2\arcmin$ from the center of NGC 1835, suggesting the exciting
possibility that it has been captured by the close massive GC and it
is on the verge of tidal disruption \citep{giusti+2023}.  In a
companion paper (Giusti et al., in preparation), we will present
the determination of the density profile and the structural parameters
of NGC 1835, as well as its dynamical age measured from the
sedimentation level of blue straggler stars following the dynamical
clock prescription \citep{ferraro+2018, ferraro+2020, ferraro+2023}.

The present work has been devoted to the discussion of the HB
morphology of NGC 1835, supported by new estimates of its distance,
reddening and age. In agreement with previous studies in the
literature, we find a distance modulus $(m-M)_0 = 18.58$
(corresponding to 52 kpc) and a color excess $E(B-V) = 0.08$.  The major result of the present
investigation is the detection of an unexpected feature: in spite of
all previous photometric studies \citep[e.g.,][]{olsen+1998}, thanks to the use of a near UV filter, our analysis
 has revealed the presence of a very extended blue tail of the HB.
Such a feature is observed in only a few cases (known as
\textit{extreme blue tailed clusters, EBTs}) in our Galaxy, including
GCs such as M13, NGC 6752, NGC 2808, NGC 2419 and M80 (see
\citealt{ferraro+1997b, ferraro+1998, dalessandro+2011, dalessandro+2013, onorato+2023}), but it had
never been observed before in any extra-Galactic cluster.

We have presented a detailed characterization of the observational
properties of the HB: {\it (i)} the optical-near UV CMDs have revealed
the presence of a pronounced blue tail extending well below the MS-TO:
the extension in magnitude is as large as 4.5 magnitudes both in the
near UV and in the optical bands; {\it (ii)} the overall colour
extension of the HB is also impressive, ranging from $(m_{\rm
  F300X}-m_{\rm F606W})\sim -2.5$ up to $(m_{\rm F300X}-m_{\rm
  F606W})\sim 2$, thus covering a wide interval of effective
temperatures, from 5000 up to 30,000 K; {\it (iii)} a large population
of RR Lyrae (67 are confirmed by OGLE, and 52 are new candidates) has
been identified; {\it (iv)} the comparison with two Milky Way GCs used
as reference has revealed that the HB of NGC 1835 looks like a
combination of the HBs of the ``classical twin GCs'' M3 and M13,
showing simultaneously the main properties of both their morphologies:
a rich population of HB stars on both sides of the instability strip
and a large sample of variable stars (as in M3), combined with an
extended blue tail (as observed in M13). 
The striking similarity with the HB morphology of M13 
suggests that also NGC 1835 should harbour a population of slowly cooling white dwarfs (see 
\citealp{chen+2021}). Unfortunately, however, their detection is beyond the photometric capacity of the current generation of instruments.
 
The HB evolutionary stage is characterized by helium combustion in the
stellar core and hydrogen burning in an adjacent shell.  The different
colour/temperature distributions of HB stars (i.e., the different HB
morphologies) observed in old stellar systems are a manifestation of
different mass distributions of the stars that reach this evolutionary
stage. More specifically, since in old stellar systems like NGC 1835
(for which we have estimated an age of $12.5\pm 1$ Gyr; see Section
\ref{sec:params}), the core mass of all HB stars is essentially set by
the occurrence of the helium flash, the key parameter responsible for
the HB morphology is the residual mass of the envelope, with large
envelope masses locating stars at red colours (i.e., low effective
temperatures), and small envelope masses placing stars at blue colors
(i.e., high $T_{\rm eff}$). One of the most obvious parameters that
can differentiate the mass of the residual envelope (hence the
distribution in colour) is the mass loss occurring along the RGB (see
\citep{origlia+2002, origlia+2007, origlia+2014}). 
However, also age spreads and helium abundance differences (possibly related to the light-element multiple population phenomenon; e.g., \citealp{milone+2014}) affect the mass of stars reaching the HB phase. Hence, understanding the origin of the HB morphology is a complex task, depending on many different parameters (see, e.g., \citealt{catelan2009, gratton+2010, milone+2014}).  While a spread of stellar ages is excluded in NGC 1835 by the apparent narrowness of its MS-TO (see Section \ref{sec:params}), the combined action of mass loss and helium spread (see e.g. \citealt{tailo+2020}) can in principle account for its complex HB morphology. Indeed, \citet{dalessandro+2013} discusses the HB morphology of M3 and M13, concluding that it can be interpreted in terms of different helium contents. Unfortunately, the photometric band combination used in this work is not optimized for detecting light-element multiple populations and, in fact, we observe no evidence of splitting or broadening along the evolutionary sequences in the CMD. A forthcoming paper will be devoted to the detailed reconstruction of the observed HB morphology through appropriate population synthesis, following an approach similar to that discussed in the cases of NGC 1904 and M13 by \citet{dalessandro+2013}.

\begin{acknowledgements}
This work is part of the project Cosmic-Lab at the Physics and
Astronomy Department ``A. Righi'' of the Bologna University (http://
www.cosmic-lab.eu/ Cosmic-Lab/Home.html). 
\end{acknowledgements}

% WARNING
%-------------------------------------------------------------------
% Please note that we have included the references to the file aa.dem in
% order to compile it, but we ask you to:
%
% - use BibTeX with the regular commands:
%   \bibliographystyle{aa} % style aa.bst
%   \bibliography{Yourfile} % your references Yourfile.bib
%
% - join the .bib files when you upload your source files
%-------------------------------------------------------------------

\end{document}